\documentclass[sigconf]{acmart}
\AtBeginDocument{%
  }
\setcopyright{acmlicensed}
\copyrightyear{2018}
\acmYear{2018}
\acmDOI{XXXXXXX.XXXXXXX}
\acmConference[Conference acronym 'XX]{Make sure to enter the correct
  conference title from your rights confirmation email}{June 03--05,
  2018}{Woodstock, NY}
\acmISBN{978-1-4503-XXXX-X/2018/06}

\usepackage{xspace}
\usepackage{subcaption} 
\usepackage{algorithm}
\usepackage{algpseudocode}

\usepackage{xcolor}

\begin{document}

\title{Multi-Faceted Large Embedding Tables for Pinterest Ads Ranking}

\author{Runze Su}
\authornote{Equal Contributions}
\email{runzesu@pinterest.com}
\affiliation{%
  \institution{Pinterest}
  \country{USA}
}

\author{Jiayin Jin}
\authornotemark[1]
\email{jjin@pinterest.com}
\affiliation{%
  \institution{Pinterest}
  \country{USA}
}

\author{Jiacheng Li}
\authornotemark[1]
\email{jiachengli@pinterest.com}
\affiliation{%
  \institution{Pinterest}
  \country{USA}
}

\author{Sihan Wang}
\authornotemark[1]
\email{sihanwang@pinterest.com}
\affiliation{%
  \institution{Pinterest}
  \country{USA}
}

\author{Guangtong Bai}
\email{gbai@pinterest.com}
\affiliation{%
  \institution{Pinterest}
  \country{USA}
}

\author{Zelun Wang}
\email{zelunwang@pinterest.com}
\affiliation{%
  \institution{Pinterest}
  \country{USA}
}

\author{Li Tang}
\email{ltang@pinterest.com}
\affiliation{%
  \institution{Pinterest}
  \country{USA}
}

\author{Yixiong Meng}
\email{ymeng@pinterest.com}
\affiliation{%
  \institution{Pinterest}
  \country{USA}
}

\author{Huasen Wu}
\email{huasenwu@pinterest.com}
\affiliation{%
  \institution{Pinterest}
  \country{USA}
}

\author{Zhimeng Pan}
\email{zpan@pinterest.com}
\affiliation{%
  \institution{Pinterest}
  \country{USA}
}

\author{Kungang Li}
\email{kungangli@pinterest.com}
\affiliation{%
  \institution{Pinterest}
  \country{USA}
}

\author{Han Sun}
\email{hsun@pinterest.com}
\affiliation{%
  \institution{Pinterest}
  \country{USA}
}

\author{Zhifang Liu}
\email{zhifangliu@pinterest.com}
\affiliation{%
  \institution{Pinterest}
  \country{USA}
}

\author{Haoyang Li}
\email{haoyangli@pinterest.com}
\affiliation{%
  \institution{Pinterest}
  \country{USA}
}

\author{Siping Ji}
\email{siping@pinterest.com}
\affiliation{%
  \institution{Pinterest}
  \country{USA}
}

\author{Degao Peng}
\email{dpeng@pinterest.com}
\affiliation{%
  \institution{Pinterest}
  \country{USA}
}

\author{Jinfeng Zhuang}
\email{jzhuang@pinterest.com}
\affiliation{%
  \institution{Pinterest}
  \country{USA}
}

\author{Ling Leng}
\email{lleng@pinterest.com}
\affiliation{%
  \institution{Pinterest}
  \country{USA}
}

\author{Prathibha Deshikachar}
\email{pdeshikachar@pinterest.com}
\affiliation{%
  \institution{Pinterest}
  \country{USA}
}

\renewcommand{\shortauthors}{Su et al.}

\begin{abstract}

Large embedding tables are indispensable in modern recommendation systems, thanks to their ability to effectively capture and memorize intricate details of interactions among diverse entities. As we explore integrating large embedding tables into Pinterest’s ads ranking models, we encountered not only common challenges such as sparsity and scalability, but also several obstacles unique to our context. Notably, our initial attempts to train large embedding tables from scratch resulted in neutral metrics. To tackle this, we introduced a novel multi-faceted pretraining scheme that incorporates multiple pretraining algorithms. This approach greatly enriched the embedding tables  and resulted in significant performance improvements. As a result, the multi-faceted large embedding tables bring great performance gain on both the Click-Through Rate (CTR) and Conversion Rate (CVR) domains. Moreover, we designed a CPU-GPU hybrid serving infrastructure to overcome GPU memory limits and elevate the scalability. This framework has been deployed in the Pinterest Ads system and achieved $1.34\%$ online CPC reduction and $2.60\%$ CTR increase with neutral end-to-end latency change.

\end{abstract}



\maketitle

\begin{figure*}[h]\label{fig:model}
  \centering
  \includegraphics[scale=0.16]{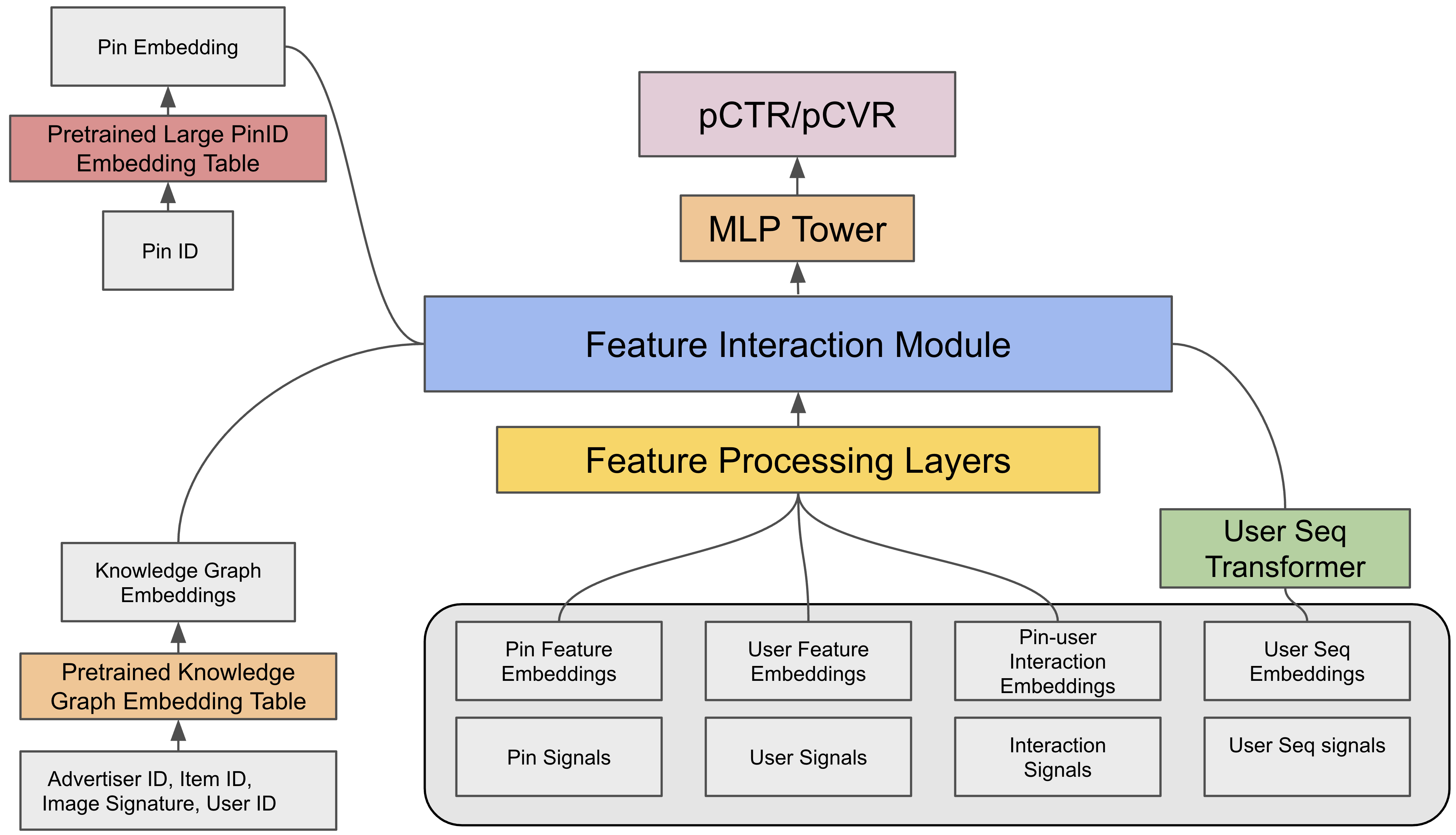}
  \caption{Large embedding table architecture of Pinterest ads models}
  \label{fig:twocolumn}
\end{figure*}

\section{Introduction}\label{sec:intro}


Advancements in modern recommender systems~\cite{zhou2018deep, song2019autoint, mao2023finalmlp, wang2021dcn, zhang2023fibinet++, wang2021masknet, zhuang2025practice}, particularly their effectiveness in predicting click-through rates (CTR) and conversion rates (CVR)~\cite{zhang2021deep, gao2023rec4ad}, have significantly accelerated the evolution of digital advertising. As a leading platform for users seeking inspiration and shopping ideas, Pinterest has been evolving and scaling up its ads ranking models to better meet advertiser goals while improving user experiences. As we scale up embedding tables, we encountered challenges unique to Pinterest's context beyond those related to model scalability. 

In recent years, Pinterest has developed several state-of-the-art, large-scale graph and sequence models for generating high-quality entity embeddings~\cite{agarwal2024omnisearchsage, pal2020pinnersage}, such as Pin embeddings (GraphSage~\cite{ying2018graph}), user embeddings (PinnerFormer~\cite{pancha2022pinnerformer}), and item embeddings (ItemSage~\cite{baltescu2022itemsage}). These embeddings encapsulate rich information regarding entity interaction histories and content attributes. Consequently, the information captured by large embedding tables~\cite{yin2024device, zhao2023embedding, qiu2025evolution} may overlap with that provided by the aforementioned pretrained embeddings. Indeed, we observed neutral offline results when training large embedding tables from scratch (see Section \ref{sec:exp}). 

To address this, we proposed a \textbf{novel multi-faceted pretraining approach} to enrich the large embedding tables with supplement information and finetune them within ads ranking models. In particular, in this work we present two pretraining methods: 1) User-Pin(item) contrastive learning (details in Section \ref{sec: pinidtable}) and 2) Large-scale heterogeneous Knowledge Graph Embedding (details in Section \ref{sec:kge}). Our approach yields significant improvements for CTR and CVR models. In particular, we observed that each pretraining method provided orthogonal gains. Further details on offline evaluation metrics can be found in Section \ref{sec:exp}.

Scaling embedding tables to tens of billions of parameters at Pinterest also presents challenges for both training and serving due to resource limitations. For training, we utilize AWS P4d instances with $8$ GPUs and a total of $320$GB GPU memory. We employed distributed model-parallel (DMP) ~\cite{ivchenko2022torchrec, zhao2023pytorch} training to shard large embedding tables across multiple GPUs. Resource limitations persist as a significant challenge in the serving phase as well. Our model serving, which operates on AWS G5.4 devices with just $24$GB GPU memory and $64$GB CPU memory per GPU, imposes even more restrictive memory constraints.

Hybrid CPU-GPU serving has been widely used in many different areas ~\cite{zhu2019graphvite, ubal2012multi2sim, xue2025hybridserve} to relax the memory limits of GPU devices. Inspired by this, we propose a scalable hybrid framework that hosts large embedding tables on external CPU clusters while maintaining the upper model on GPUs. This design enables scaling embedding tables independently of GPU capacity. Additionally, we apply post-training INT4 quantization to compress embedding tables to approximately $40\%$ of their original Float16 size, further reducing CPU memory and cost.

While this hybrid approach introduced communication overhead between CPU-based embedding fetching and GPU inference, we mitigated potential serving latency increases by initiating CPU embedding fetching as early as possible, allowing it to run in parallel with other serving components, which led to neutral overall latency. Furthermore, we implemented robust version synchronization to ensure consistency between CPU and GPU models throughout deployment and serving. Additional details are provided in Section \ref{sec:serving}.

\begin{figure*}

  \centering
  \includegraphics[scale=0.13]{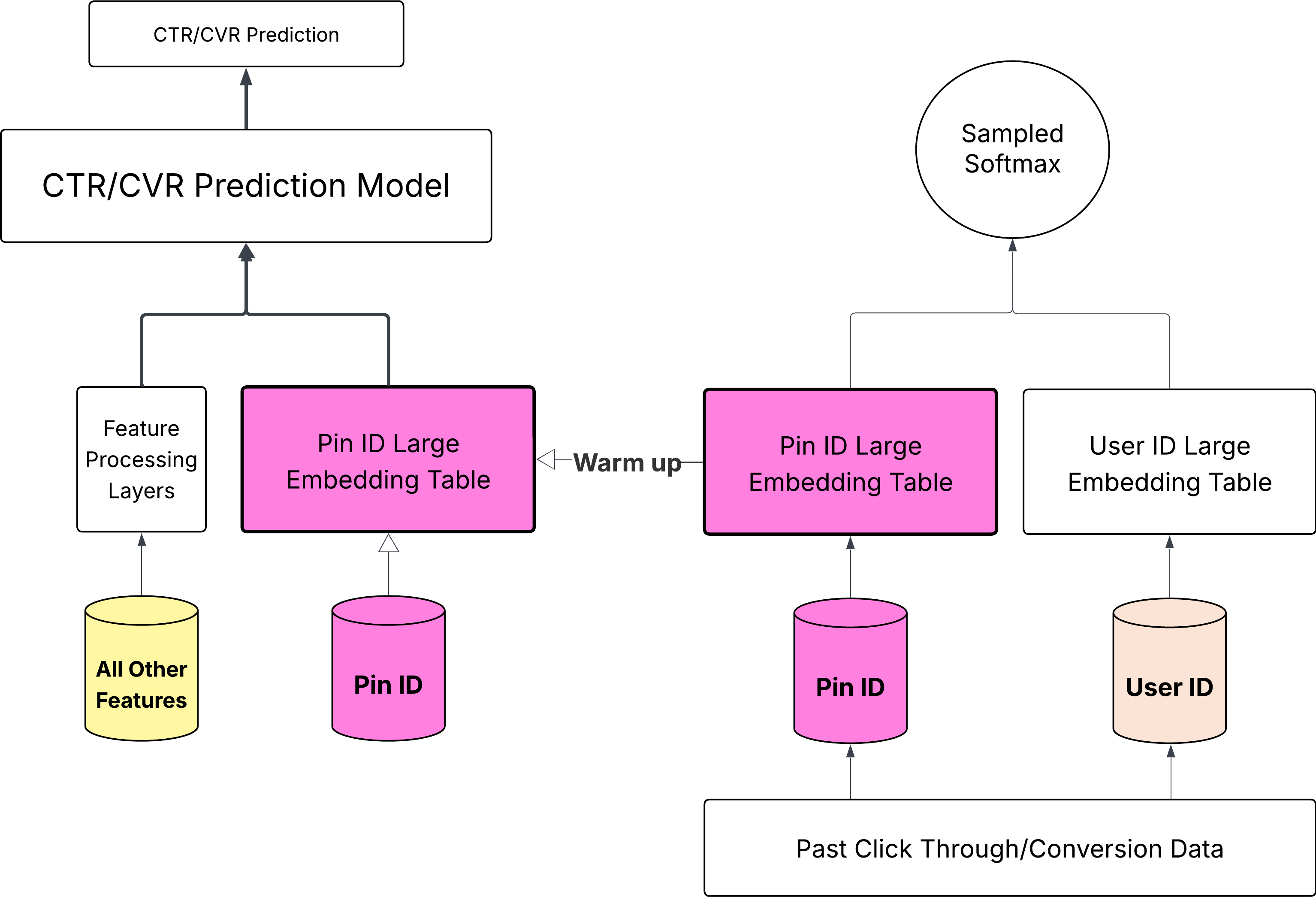}
  \caption{Pretrained Pin ID Embedding Table}
  \label{fig:pretrained_atg}
\end{figure*}

\begin{figure*}[h]
  \centering
  \includegraphics[scale=0.35]{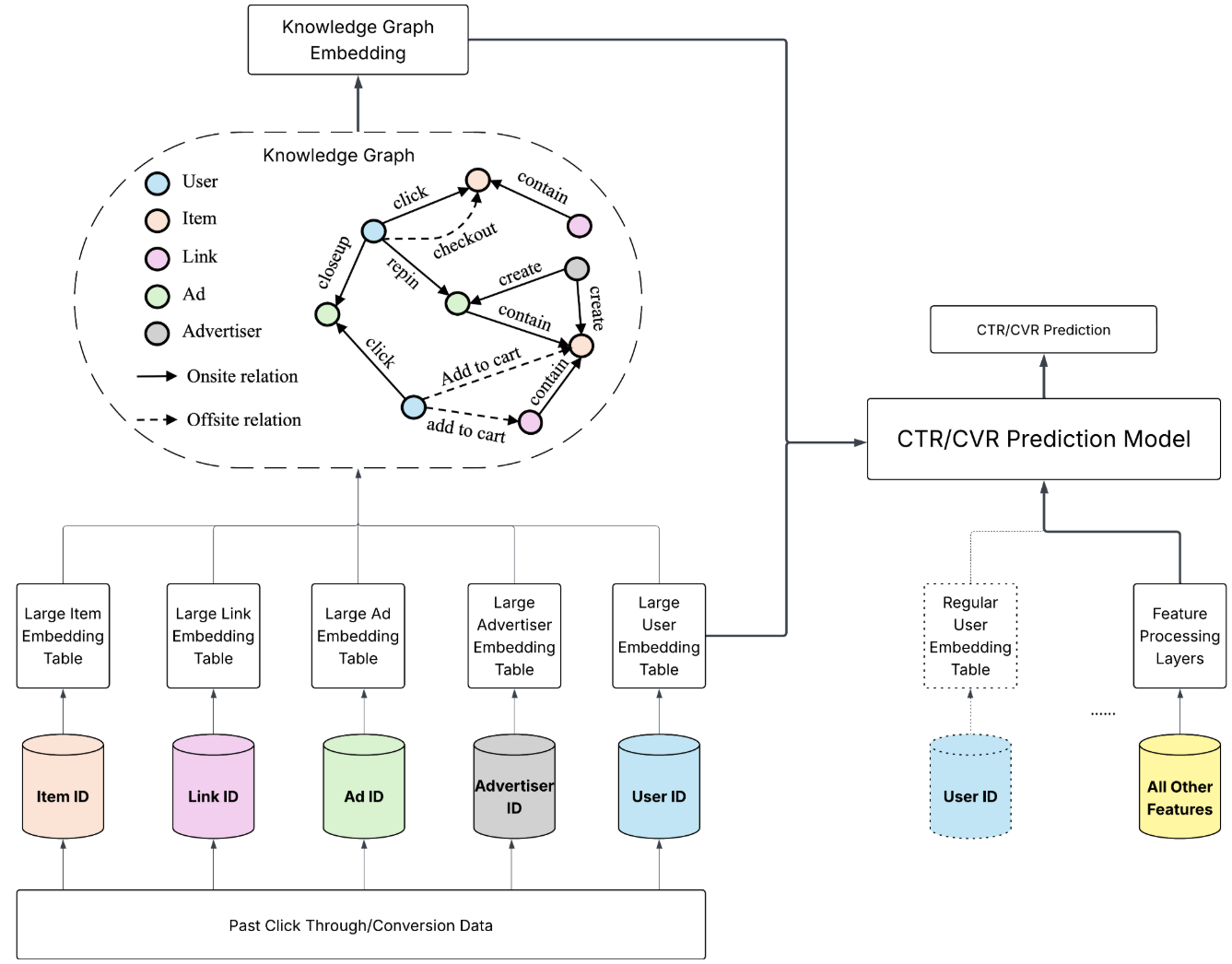}
  \caption{Pretrained Knowledge Graph Embedding}
  \label{fig:pretrained_kge}
\end{figure*}

\section{Problem Setup and Background}\label{sec:problem}
\subsection{Ads Ranking Models}
Pinterest Ads system provides personalized recommendations of Advertisement Pins to hundreds millions of users. The efficiency of ads recommendation system replies heavily on the predicted click-through rate (CTR) and the conversion rate (CVR) of ranking models. The click through rate is the probability of the occurrence of a click given an ad impression:
\begin{equation*}
    CTR = P(\text{click}\mid \text{impression}).
\end{equation*}
There two types of conversion rate, click-based (click-through) , the probability of conversion after a user clicks a Pin: $$CCVR = P(\text{conv} \mid \text{click});$$ and impression-based (view-through) conversions, the probability of conversion after a user view an impression without clicking: $$P(\text{conv} \mid \text{impression}, \text{no click}).$$ In the auction state, the utility is a combination of these two types of CVR:
\begin{equation*}
    P(\text{conv}) = CTR \cdot P(\text{conv} \mid \text{click}) + P(\text{conv} \mid \text{impression}, \text{no click}).
\end{equation*}

Facilitated by the development of Deep Learning, modern ads ranking models are capable of extracting hidden information from thousands of input features and make accurate predictions ~\cite{li2023deep, liu2022review}. Roughly speaking, Pinterest ads ranking models utilize three types of features, including 
\begin{itemize}
\item Pin(item) features denoted by $P = \big\{P_1, P_2, ..., P_{N_1} \big\}$, such as Prior CTR, CVR of a Pin aggregated over some time window, content embeddings, etc. 
\item User features denoted by $U = \big\{U_1, U_2, ..., U_{N_2}\big\}$, including counting features such as user spend, demographic features, user embeddings and sequence features built based users' past engagement and conversion history. 
\item User-Pin interaction features denoted by $I = \big\{I_1, I_2, ..., I_{N_3} \big\}$. For example, user-advertiser interaction counts etc.  
\end{itemize}

Given a user and a Pin, our ads ranking system fetches the relevant features and inputs them into either a CTR or CVR model, depending on the ad type, to generate predictions for CTR or CVR values (specifically, CCVR and VTCVR). The CTR prediction model can be represented as a function $$f_{CTR}(P, U, I) \rightarrow [0, 1]$$ and the CVR prediction model can be represented as $$f_{CVR}(P, U, I) \rightarrow [0, 1]\times[0,1].$$ 

\subsection{Large Embedding Tables}\label{sec:large_emb}
Embedding tables representing high-dimensional objects as low-dimensional vectors are fundamental to modern recommendation systems. In our architecture, high-cardinality categorical features are mapped to dense embeddings via hashed lookup tables~\cite{kang2021learning, liu2022monolith}. A key challenge of this approach is managing hash collisions, which occur when the number of unique identifiers exceeds the embedding table’s predefined size. A straightforward yet practical mitigation is to increase the table’s vocabulary size~\cite{yusuf2021collision}; however, this introduces significant challenges, including issues related to scalability in training and serving, as well as increased sparsity, as discussed in Section~\ref{sec:intro}~\cite{zhang2022towards, hsu2024taming}. In the remainder of the paper, we present our strategies for addressing the challenges associated with scaling embedding tables in Pinterest Ads Ranking models.

\section{Methodology for Scaling up Embedding Tables}
\subsection{Challenges and Strategies}\label{sec:scale_table}
As mentioned in Section \ref{sec:large_emb}, large embedding tables are a efficient technique to reduce hash collisions for high cardinality id features. However, when we initially incorporated this technique into our ads ranking models for high cardinality IDs, such as Pin ID, User ID, Image Signature ID, and Item ID, it did not yield any performance gain. This indicates that without further treatment, these large embedding tables are not able to capture extra information for our models. 

Figure \ref{fig:twocolumn} illustrates our strategy to overcome this challenge. The main idea is to enrich the embedding tables by various pretraining methods, and followed by fine-tuning pretrained embeddings within our ads ranking models. Further details are provided in Section~\ref{sec:pretrained_tables}..

Our large embedding tables currently contain approximately $450$ million rows and do not fit into a single GPU. As the business continue to grow, the size of these embedding tables is expected to scale up in future iterations. To address this limitation, we employ the TorchRec library to shard the large embedding tables across 8 different GPUs during training. For serving, we implemented a scalable CPU-GPU hybrid serving pipeline, as detailed in Section~\ref{sec:serving}.

\begin{figure*}[h]
  \centering
  \includegraphics[scale=0.4]{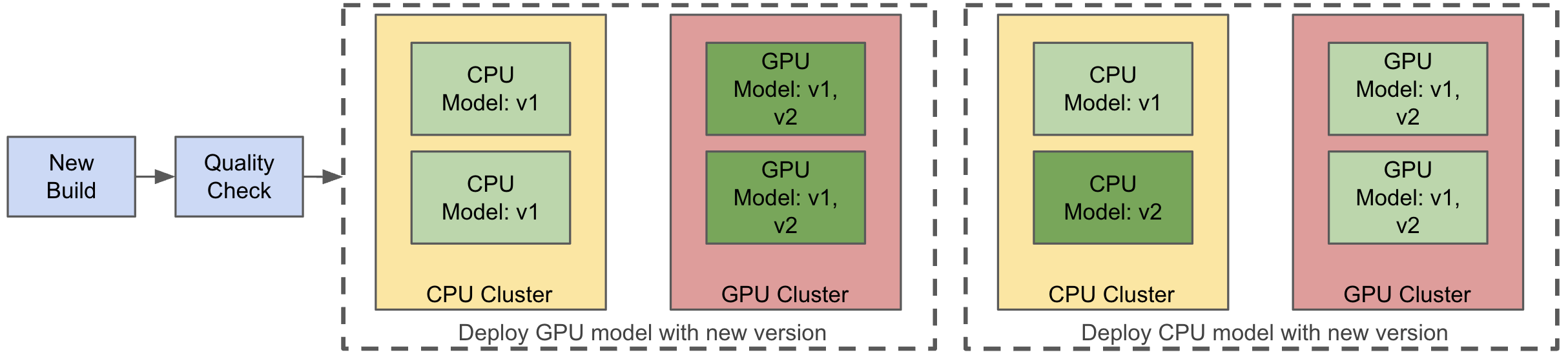}
  \caption{Model Deployment Process of Hybrid Serving}
  \label{fig:deployment}
\end{figure*}

\begin{figure}[h]
  \centering
  \includegraphics[scale=0.12]{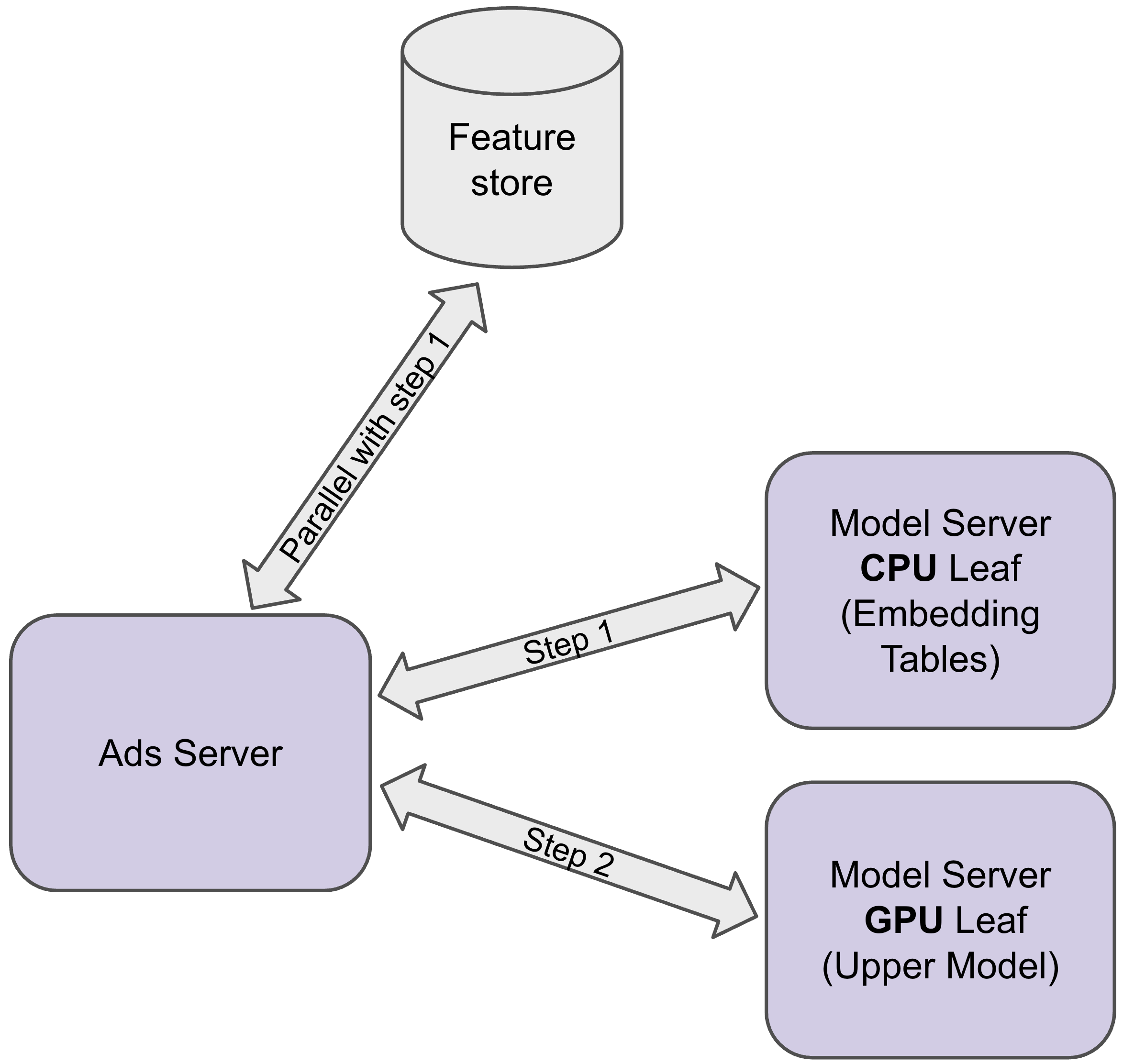}
  \caption{Hybrid Serving Pipeline}
  \label{fig:serving}
\end{figure}

\subsection{Pretrained Large Embedding Tables}\label{sec:pretrained_tables}

\subsubsection{User-Pin Contrastive Learning}\label{sec: pinidtable}
We first adopt the contrastive learning pretraining method introduced in \cite{hsu2024taming}. Figure \ref{fig:pretrained_atg} shows the core objectives of this approach: (1) to independently capture interactions between users and Pins without interference from other features; and (2) to leverage richer data during pretraining, thereby enabling the embedding tables to contain information from longer history. \footnote{Use of these signals is subject to applicable user privacy choices.} More specifically, we extract historical interactions between users and Pins from both onsite engagement and conversion data, and use these interactions to pretrain the user and Pin embedding tables with a contrastive loss. 
This approach involves constructing two large embedding tables for both Pin and user IDs respectively, and utilizes a large amount of historical click-through and conversion data for pretraining. We applied contrastive loss during pretraining, with both in-batch negatives and randomly sampled out-batch negatives. 

\subsubsection{Heterogeneous Knowledge Graph Embedding Table}\label{sec:kge}
In addition to the contrastive learning tables, we further introduced pretrained large-scale Knowledge Graph Embedding (KGE) tables. As illustrated in Fig. \ref{fig:pretrained_kge}, the heterogeneous graph incorporates both onsite engagement edges and opt-in offsite conversion edges. Node entity embeddings are trained via a link prediction task, which predicts the existence of an edge between a given pair of head and tail nodes. This graph consists of billions of nodes and edges, significantly larger in scale than production ranking models, thus providing substantially supplementary information. Moreover, pretrained embeddings at Pinterest, such as GraphSage\cite{ying2018graph} and ItemSage\cite{baltescu2022itemsage}, are based on Graph Convolutional Networks (GCNs). In contrast, KGE models~\cite{TransE, TransR} follow a distinct modeling approach, potentially capturing additional information that may be overlooked by GCN-based methods.

\subsection{CPU-GPU Hybrid Serving}\label{sec:serving}
Scaling the serving of large embedding tables poses significant challenges. Existing solutions, such as Monolith\cite{liu2022monolith}, utilize stand-alone parameter servers for large embedding tables. However, those methods don't explicitly guarantee version synchronization between large embedding tables and upper-level GPU models, due to its online learning nature.

In this section, we present an effective hybrid serving solution tailored for Pinterest Ads ranking system which guarantees version synchronization without introducing extra serving latency. Currently, Pinterest Ads ranking models are hosted on g5.4xlarge instances, each limited to $24$GB of GPU memory and $64$GB of CPU memory. Therefore, it is impractical to store large embedding tables in GPU memory. Moreover, hosting the embedding tables in the CPU memory of the same host as the upper ranking model does not scale efficiently. A commonly adopted solution is to treat embeddings as precomputed features and upload them to a feature store. However, this approach creates substantial difficulties in ensuring version consistency between the embeddings and the upper model, heightening the risk of version mismatches and potential performance degradation. To address these challenges, we propose a CPU-GPU hybrid serving pipeline, which provides several key advantages:
\begin{itemize}
    \item Guarantee the versions between the CPU modules and GPU modules;
    \item Better scalability of large embedding tables; 
    \item Low infrastructure cost and serving latency.
\end{itemize}


\subsubsection{Hybrid Model Deployment}\label{sec:deploy}
Our hybrid deployment process is illustrated in Figure ~\ref{fig:deployment} and described in Algorithm \ref{alg:deployment}. In particular, the process in Phase 2 ensures \textbf{version consistency} between the CPU-based embedding models and the GPU-based upper models during the transition stage of model deployment.
\begin{algorithm}[H]
\caption{Version-Consistent Deployment Protocol}
\label{alg:deployment}
\begin{algorithmic}
\Require Stable models: $M_{CPU}^{old}$, $M_{GPU}^{old}$
\Require Candidate models: $M_{CPU}^{new}$, $M_{GPU}^{new}$
\Ensure Transition to $M_{CPU}^{new}$, $M_{GPU}^{new}$ without version conflict

\State \textbf{Phase 1: Deploy New GPU Model}
\State Deploy $M_{GPU}^{new}$ alongside $M_{GPU}^{old}$\;
\State \hspace{1em} ($M_{GPU}^{new}$ is inactive; all traffic uses $M_{CPU}^{old} \rightarrow M_{GPU}^{old}$)

\State

\State \textbf{Phase 2: Deploy New CPU Model and Transition}
\State Deploy $M_{CPU}^{new}$ to embedding service\;
\While{$M_{CPU}^{old}$ is not decommissioned}
    \State Receive inference request
    \State $(E, V) \gets \text{GenerateEmbeddings}(\text{request})$ \Comment{$E$: embeddings, $V$: version\_id}
    \If{$V$ is $M_{CPU}^{new}$}
        \State $\text{score} \gets \text{Score}(E, M_{GPU}^{new})$
    \Else
        \State $\text{score} \gets \text{Score}(E, M_{GPU}^{old})$
    \EndIf
\EndWhile

\State

\State \textbf{Phase 3: Deprecate Old Models}
\State Decommission $M_{CPU}^{old}$ and $M_{GPU}^{old}$
\State \Return System runs on $M_{CPU}^{new} \rightarrow M_{GPU}^{new}$

\end{algorithmic}
\end{algorithm}

\subsubsection{Hybrid Model Serving Process}\label{serve}
Our CPU-GPU hybrid serving pipeline is illustrated in Figure~\ref{fig:serving}. The end-to-end inference process for a given request is detailed in Algorithm \ref{alg:serving}. In particular, this two-phase remote procedure call (RPC) structure guarantees that an embedding feature is always scored by its corresponding model version.

\begin{algorithm}[H]
\caption{Hybrid CPU-GPU Serving Pipeline}
\label{alg:serving}
\begin{algorithmic}
\Require Incoming request ($\texttt{req}$) at the Ads Server ($S_{Ads}$)
\Ensure Final $\texttt{score}$ computed with version-matched models

\Statex
\textbf{Actors:}
\Statex \hspace{\algorithmicindent} $S_{Ads}$: Ads Server
\Statex \hspace{\algorithmicindent} $S_{CPU}$: Model Server CPU Leaf (embedding model)
\Statex \hspace{\algorithmicindent} $S_{GPU}$: Model Server GPU Leaf (upper model)
\Statex

\Procedure{HandleInferenceRequest}{req}
    \Statex \textbf{Step 1: Embedding Retrieval} 
    \State Initiate RPC from $S_{Ads}$ to $S_{CPU}$ to retrieve embeddings for $\texttt{req}$
    \State $(E, V) \gets S_{CPU}.\text{GenerateEmbeddings}(\texttt{req})$ \Comment{$E$: embeddings, $V$: version ID}
    \State $S_{Ads}$ receives $(E, V)$ from $S_{CPU}$

    \Statex \textbf{Step 2: Version-Aware Scoring}
    \State $S_{Ads}$ constructs a scoring request using $E, V$
    \State Initiate RPC from $S_{Ads}$ to $S_{GPU}$ for scoring
    \State $\texttt{score} \gets S_{GPU}.\text{ScoreWithMatchingModel}(E, V)$
    \State $S_{Ads}$ receives $\texttt{score}$ from $S_{GPU}$

    \State \Return $\texttt{score}$
\EndProcedure
\end{algorithmic}
\end{algorithm}

\subsection{Serving Parallelism Optimization}\label{serve}
Embedding lookups and CPU-GPU communication can introduce significant latency during serving. To mitigate this, we optimized our hybrid serving pipeline to initiate embedding lookup and transfer as early as possible in the process. As a result, we achieve no increase in end-to-end serving latency.

\begin{table*}[h]
    \centering
    \begin{tabular}{l c c c c}
        \toprule
        \textbf{Engagement/Conversion} & \textbf{Data Domain} & \textbf{Use Pretrained Table} & \textbf{ROC\_AUC}\\
        \midrule
        CTR  & Click-through data &  & $+0.01\%$\\
        CTR & Click-through data & \checkmark & $\mathbf{+0.09\%}$ \\
        Click Checkout CVR & Conversion data & & $+0.02\%$ \\
        Click Checkout CVR & Conversion data &\checkmark & $\mathbf{+0.16\%}$\\
        View Checkout CVR & Conversion data & & $+0.02\%$ \\
        View Checkout CVR & Conversion data &\checkmark & $\mathbf{+0.11\%}$ \\
        \bottomrule
    \end{tabular}
    \caption{Offline performance gains for CTR and CVR models using large embedding tables.}
    \label{tab:results}
\end{table*}

\section{Experiments and Results}\label{sec:exp}
\subsection{Offline Experiments}

\subsubsection{Multi-facet Pretraining}

We performed offline experiments on both the CTR and CVR prediction models. The best-performing configurations are as follows:
\begin{itemize}
    \item \textbf{Click-through Rate Prediction Model:} We utilize a large Pin ID embedding table alongside a large knowledge graph embedding table, incorporating advertiser IDs, image signature IDs, and item IDs.
    \item \textbf{Checkout Conversion Prediction Model:} We employ a large knowledge graph embedding table with advertiser IDs, image signature IDs, item IDs, and user IDs.
\end{itemize}

Offline experimental results, as shown in Table \ref{tab:results}, demonstrated the efficacy of the pretraining methodology compared to the training-from-scratch approach. On both the CTR and Checkout CVR datasets, our pre-training technique delivered a relative performance lift that was more than four-fold greater. 

Furthermore, a subsequent ablation study conducted on the CTR model indicated that our two pretraining strategies (described in Sections 3.2.1 and 3.2.2) provided orthogonal improvements: contrastive learning on Pin IDs first contributed a +0.03\% gain in ROC AUC, and the Knowledge Graph Embedding (KGE) framework then provided an additional +0.06\% improvement on top of that. 

\subsubsection{Model Quantization}
Our large embedding tables remain substantial in size even with half-precision and after TorchScript serialization. To further improve efficiency and scalability, we employed Post-Training Quantization (PTQ) technique to compress large embedding tables. By performing INT4 quantization, the embedding table size was reduced by $60\%$. Our offline evaluation on the CVR model yielded that with quantized large embedding table, the model not only matched but slightly exceeded the performance of large embedding tables with half precision, registering a $+0.03\%$ gain in AUC. It should be noticed that this result is consistent with findings in prior,  such as ~\cite{zhou2024dqrm}. A leading hypothesis is that the digit loss in quantization acts as a form of regularization. This mitigates the overfitting on the sparse and potentially noisy parameters typical of large-scale embedding tables.

\subsection{Online Experiments}
We tested the large embedding table performance on the RP surface of Pinterest Click through rate prediction. Table~\ref{tab:online_results} shows the online performance on large embedding tables. Beside the general cost per click and click-through rate, we also defined two online metrics for to measure the quality of online click through rates:
\begin{itemize}
    \item \textbf{Good click-through rate:} The proportion of clicks where the session duration exceeds 30 seconds.
    \item \textbf{Outbound click-through rate:} The proportion of clicks leading to external websites or landing pages.
\end{itemize}

\begin{table}[h]
    \centering
    \begin{tabular}{l c}
        \toprule
        \textbf{Online CTR Metrics} & \\
        \midrule
        CPC Ads CPC & -1.34\%\\
        Ads Clicks Per Dollar & +1.89\%\\
        Platform wise CTR & +2.60\%\\
        Platform wise gCTR & +3.52\%\\
        Platform wise oCTR & +2.66\%\\
        \bottomrule
    \end{tabular}
    \caption{Online performance for CTR model using large embedding tables.}
    \label{tab:online_results}
\end{table}

The large embedding table CTR model demonstrates significant improvements across all core online metrics. Furthermore, the model introduced zero serving latency increase and a negligible rise in serving cost, thanks to the hybrid serving infrastructure (Section \ref{sec:serving}).

\section{Ablation Studies}

\textbf{Should we freeze the pretrained embedding tables?} We tested two different approaches to integrate pretrained embedding tables into the downstream model: 1. freeze the pretrained large embedding table and 2. fine-tune the pretrained large embedding table with downstream CTR/CVR tasks. From our offline experiment, we observed $-0.01\%$ ROC\_AUC if we freeze the pretrained large embedding table, while fine-tuning it could bring $+0.09\%$ ROC\_AUC. This demonstrates the necessity of fine-tuning pretrained embedding tables in downstream tasks. 

\textbf{Pretrained embedding table staleness.} 
We also studied how the staleness of pretrained embedding table would impact the offline performance gain on CTR models. The staleness here means the training data time gap between pretraining and downstream fine-tuning. For example, if we use data from January 1st to March 31st to perform pretraining and use data from April 1st to June 30th to perform downstream fine-tuning, there is no staleness; if we use data from October previous year to January 1st to perform pretraining and still use data from April 1st to June 30th for downstream fine-tuning, then there is a 3-month staleness. In table \ref{tab:staleness_results}, we summarize the experimental results. We can see that the 3-month staleness leads to -0.05\% ROC\_AUC decay and the 6-month staleness completely eliminates the gain from pretrained large embedding tables. 
 
\begin{table}[h]
    \centering
    \begin{tabular}{l c}
        \toprule
        Staleness & \textbf{ROC\_AUC}  \\
        \midrule
        No Staleness & +0.09\% \\
        3 months staleness & +0.04\% \\ 
        6 months staleness & +0.00\% \\
        \bottomrule
    \end{tabular}
    \caption{Offline performance gain for CTR model under different level of staleness.}
    \label{tab:staleness_results}
\end{table}

\section{Conclusion and Future Work}
In summary, our work provides a foundation for robust, scalable, and high-performing ad recommendation systems capable of handling the ever-increasing data volume and complexity of commercial platforms like Pinterest. While our framework offers significant accuracy improvements, the use of large embedding tables presents opportunities for optimization. One future work could be the shared embedding tables for ID processing in user sequences. Other promising directions include exploring more efficient sharding and caching strategies, hierarchical or adaptive embedding structures, and fine-tuning refresh policies for embedding versions in the serving stack to further minimize latency and maintain strict version synchronization during rapid iteration.

\section*{Acknowledgements}
We would like to express our sincere gratitude to all those who contributed to the development of our online recommendation model. We especially thank Paulo Soares, Degao Peng, Qifei Shen, Jiankai Sun, Dontao Liu, and Jinfeng Zhuang for their work on the Engagement models; Andrew Qiu, Shubham Barhate, Rafael Müller, Sourav Bhattacharjee, and Shayan Ehsani for their contributions to the Conversion models; Chuizheng Meng, Jiarui Feng, Chongyuan Xiang, and Yang Tang for KGE feature development. Our thanks also go to Yuan Wang, Longyu Zhao, and Peifeng Yin for sharing their experience on retrieval models, as well as Nan Li, Lida Li, Cindy Chen, and Archer Liu for their general support. Their expertise, dedication, and collaboration were essential to the success of this project.

\bibliographystyle{ACM-Reference-Format}
\bibliography{large_embed_bib}

\end{document}